\documentclass{revtex4}
\usepackage{amssymb}
\usepackage{graphicx}
\begin{document}
\title{ Resonance spectroscopy of  gravitational states of antihydrogen}
\author{A. Yu. Voronin$^1$, V.V. Nesvizhevsky$^2$, O.D. Dalkarov$^1$, E.A. Kupriyanova$^1$, P.Froelich$^3$}
\affiliation{ $^1$ P.N. Lebedev Physical Institute, 53 Leninsky
prospect, 117924 Moscow, Russia.
\\
$^2$ Institut Laue-Langevin (ILL), 6 rue Jules Horowitz,
 F-38042, Grenoble, France.
\\
$^3$Department of Quantum Chemistry, Uppsala University, Box 518,
SE-75120 Uppsala, Sweden.
}

\begin{abstract}

We study a method to induce resonant transitions between antihydrogen ($\bar{H}$) quantum states above a material surface in the gravitational field of the Earth. The method consists in applying a gradient of magnetic field which is temporally oscillating with the frequency equal to a frequency of a transition between gravitational states of antihydrogen. Corresponding resonant change in a spatial density of antihydrogen atoms  can be measured as a function of  the frequency of  applied field. We estimate an accuracy of measuring antihydrogen gravitational states  spacing and show how a value of the gravitational mass of the $\bar{H}$ atom can be deduced from such a measurement.

\end{abstract}

\maketitle

\section{Introduction}
Precision tests of the Equivalence Principle (EP) in different physical phenomena is of fundamental interest. This statement is especially valid for testing EP in case of a quantum motion of  antiatoms. Detailed studies of gravitational properties of antimatter are planned by most  groups involved in experiments with antihydrogen ($\bar{H}$) \cite{Alpha,AlphaGrav, Yam, gabr10,  cesa05, Aegis1}.

Here we discuss a possibility of exploring  gravitational states of $\bar{H}$ \cite{GravStates} with potentially very precise spectroscopic methods. We study the behaviour of $\bar{H}$ bounded in the gravitational field near a material surface under the influence of alternating magnetic field, with the frequency adjusted to induce resonance transitions between lowest gravitational states. This approach gives access to the gravitational energy level spacing and thus to the gravitational mass of $\bar{H}$.

\section{Gravitationally bound Antihydrogen in alternating magnetic field}
%In the following we will study transitions between  gravitational states of $\bar{H}$ center of mass (c.m), which are induced by an alternating magnetic field.
In this section we will study a motion of an $\bar{H}$ atom, localized in the gravitational state near a horizontal plane mirror, under the influence of the alternating magnetic field.

The interaction of magnetic field  with \emph{a moving through the field } ground state $H$ atom  \cite{LambH,GorkH,LozH} is dominated by interaction of an average  magnetic moment of an atom \cite{LL} in given hyperfine state with the magnetic field.  We will focus on the alternating  magnetic field with   a gradient in the vertical direction. This condition is needed for coupling  the field and the center of mass (c.m) $\bar{H}$ motion in the gravitational field of the Earth. It allows one to induce resonant transitions between quantum gravitational states of $\bar{H}$ \cite{GravStates}. Such states have similar properties with those, discovered for neutrons \cite{Nature1,nesv00,nesv03,NVP,EPJC,NeutrWaveGuide}, in particular they are characterised by the following  energy ($\varepsilon_n$) and spatial ($H_n$) values:
\begin{eqnarray}\label{eGrav}
&\varepsilon_n&=\sqrt[3]{\frac{\hbar^2M^2g^2}{2m}} \left(\lambda_n+\Delta \right),\\
&H_n&=\sqrt[3]{\frac{\hbar^{2}}{2mMg}} \lambda_n, \label {Hn} \\
&\mathop{\rm Ai}(&-\lambda_n)=0.
\end{eqnarray}
 Here $M$ is the gravitational mass of $\bar{H}$, $m$ is the inertial mass of $\bar{H}$ (we distinguish between $M$ and $m$ in view of discussing EP test), $g$ is a gravitational field intensity near the Earth surface, $\mathop{\rm Ai}(x)$ is an Airy function \cite{abra72}, $\Delta\simeq-i0.005$ is a universal complex shift of gravitational levels due to the interaction with a material (perfectly conducting) wall \cite{GravStates}. All the states acquire equal width depending on material surface substance $\Gamma=2\sqrt[3]{\frac{\hbar^2M^2g^2}{2m}}|\Delta|$. This width  corresponds to the lifetime of $0.1$ s in case of a perfectly conducting surface and is twice larger for silica \cite{voro05,QRslabs,QRnanopor}.

   % The problem of an atom motion in nonhomogeneous alternating magnetic field is complicated by coupling of the center of mass (c.m.) motion with interatomic degrees of freedom including spin.
    %However, we will show that for the field intensities and characteristic frequencies of interest the adiabatic approximation is justified, which means that internal state of an atom "follows" at any point in space and at any %instant of time the value of magnetic field $\vec{B}(\vec{r},t)$.

   % The problem of separation of the center of mass motion and relative motion in case of hydrogen atom  in homogeneous magnetic field  was solved in %\cite{Lamb,Dzyal}. The problem we are interested in here is more complicated, % namely antihydrogen atom bounces in the gravitational field of the %Earth above material surface superposed with \emph{nonhomogeneous} time-varying magnetic field $\vec{B}$.
    We will consider  the magnetic field in the form:
% \begin{equation}\label{Magn}
%\vec{B}(z,x,t)=\left(B_0+ B_z(z,x,t)\right) \vec{e}_z+B_x(z,x,t) \vec{e}_x
%\end{equation}
\begin{equation}\label{Magn1}
\vec{B}(z,x,t)=B_0 \vec{e}_z+ \beta \cos(\omega t) \left(z \vec{e}_z-x \vec{e}_x \right).
\end{equation}

Here $B_0$ is an amplitude of a constant, vertically aligned,  component of magnetic field, $\beta$ is the value of magnetic field gradient, $z$ is a distance measured in a vertical direction, $x$ is a distance measured in horizontal direction, parallel to a surface of a mirror.

Time-varying magnetic field (\ref{Magn1}) is accompanied with an electric field ($[\vec{\nabla} \vec{E}]=-\frac{1}{c} \partial \vec{B}/\partial t$). However, for the ultracold atom velocities, corresponding interaction terms are small  and thus will be omitted.

Inhomogeneous magnetic field couples the spin and spatial degrees of freedom. $\bar{H}$ wave-function in this case is described by a four-component column (in a nonrelativistic treatise) in the spin space, each component being a function of the c.m. coordinate $\vec{R}$, relative $\bar{p}-\bar{e}$ coordinate $\vec{\rho}$ and time $t$.
The corresponding Schr\"{o}dinger equation is:
\begin{equation} \label{Schr}
i \hbar\frac{\partial \Phi_{\alpha}(\vec{R},\vec{\rho},t)}{\partial t}=\sum_{\alpha'}\left[ -\frac{\hbar^2}{2m}\Delta_R+Mgz+V_{CP}(z) +\widehat{H}_{in}+\widehat{H}_m \right]_{\alpha, \alpha'} \Phi_{\alpha'}(\vec{R},\vec{\rho},t).
\end{equation}
Subscript $\alpha$ indicates one of four spin states of $\bar{p}-\bar{e}$ system.
The meaning of the interaction terms is the following.

$V_{CP}(z)$ is a atom-mirror interaction potential, which at asymptotic atom-mirror distance turns into the van der Waals/ Casimir-Polder potential  (see \cite{voro05,voro05l} and references there in).
$\widehat{H}_{in}$ is the Hamiltonian of interatomic motion, which includes hyperfine interaction.
\begin{equation}\label{Hc}
\widehat{H}_{in}=-\frac{\hbar^2}{2 \mu}\Delta_\rho-e^2/\rho+\frac{\alpha_{HF}}{2}\left(\hat{F}^2-3/2\right).
\end{equation}
Here $\mu=m_1m_2/m$, $m_1$ is the antiproton mass, $m_2$ is the positron mass, $m=m_1+m_2$, $\alpha_{HF}$ is the hyperfine constant, $\hat{F}$ is the operator of the total spin  of antiproton and positron. We will treat only $\bar{H}$ atoms in a $1S$-state (below we will show  that excitation of other states in a studied process is improbable). The  term $\frac{\alpha_{HF}}{2}\left(\hat{F}^2-3/2\right)$ is a model operator, which effectively accounts for the hyperfine interaction and  reproduces the hyperfine energy splitting correctly.

The term $\widehat{H}_m$ describes the field-magnetic moment interaction:
\begin{equation}\label{Hm}
\widehat{H}_m=-2\vec{B}(z,x,t)\left( \mu_{\bar{e}}\hat{s}_{\bar{e}}+ \mu_{\bar{p}}\hat{s}_{\bar{p}}\right).
\end{equation}
Here $\mu_{\bar{e}}$ and $\mu_{\bar{p}}$ are magnetic moments of positron and antiproton respectively.

As far as the field $\vec{B}(z,x,t)$ changes spatially and temporally, this term couples the spin and the c.m. motion.

We will assume that in typical conditions of a spectroscopy experiment $\bar{H}$  velocity $v$ parallel to the mirror surface ( directed along $x$-axis) is of order of a few $m/s$ and is much larger than a typical vertical velocity in lowest gravitational states (which is of order of $cm/s$). We will treat motion in a frame moving with the velocity $v$ of the $\bar{H}$ atom along the mirror surface.  Thus we will consider the x-component motion as a classical motion with a given velocity $v$ and will substitute $x$-dependence by $t$-dependence. We will also assume that $B_0\gg \beta L$, where $L\sim 30$ cm is a typical size of an experimental installation of interest . This condition is needed for "freezing" the magnetic moment of an atom along the vertical direction; it provides the maximum transition probability.

We will be interested in the weak field case, such that the Zeeman splitting is much smaller than the hyperfine level spacing $\mu_B B_0\ll \alpha_{HF}$. The hierarchy of all mentioned above interaction terms  can be formulated as follows:
\begin{equation}\label{hierar}
m_2e^2/\hbar^2 \gg \alpha_{HF}\gg \mu_{\bar{e}} |B_0 |\gg \varepsilon_n,
\end{equation}
and thus it assumes  using the adiabatic expansion for solving  Eq.(\ref{Schr}); it is based on the fact that an internal state of an $\bar{H}$ atom follows adiabatically the spatial and temporal variations of external magnetic field.  Neglecting non-adiabatic coupling,  an equation system for the amplitude  $C_n(t)$ of a gravitational state $g_n(z)$ has the form:
\begin{equation}\label{Adiab}
i \hbar \frac{d C_{n}(t)}{dt}= \sum_{k} C_{k}(t) V(t)_{n,k}\exp \left(-i\omega_{n k} t\right ).
\end{equation}
Transition frequency   $\omega_{n k}=(\varepsilon_k-\varepsilon_n)/\hbar$ is determined by gravitational energy level spacing. This fact is used in the proposed approach to access gravitational level spacing by scanning the applied field frequency, as will be explained in the following.

In this formalism the role of the coupling potential $V(z,t)$ is played by the  energy of an atom in a fixed hyperfine state thought of as a function of (slowly varying) distance $z$ and time $t$.
\begin{equation}
V(t)_{n,k}=\int_0^\infty g_n(z) g_{k}(z) E(t,z)dz.\\
\end{equation}
The gravitational state wave-function is given by the Airy function \cite{GravStates}.

Energy $E(z,t)$ is the eigenvalue of internal and magnetic interactions $\widehat{H}_{in}+\widehat{H}_m$, where the c.m. coordinate $\vec{R}$ and time $t$ are treated as slow-changing parameters.
Corresponding expressions for   eigen-energies of a $1S$ manifold are:

\begin{eqnarray}\label{Ea}
E_{a,c}&=&E_{1s}-\frac{\alpha_{HF}}{4}\mp\frac{1}{2}\sqrt{\alpha_{HF}^2+|(\mu_B-\mu_{\bar{p}})B(z,t)|^2},  \\ \label{Eb}
E_{b,d}&=&E_{1s}+\frac{\alpha_{HF}}{4}\mp \frac{1}{2}|(\mu_B+\mu_{\bar{p}})B(z,t)|.
\end{eqnarray}

Subscripts $a,b,c,d$ are standard notations for  hyperfine  states of a $1S$ manifold in the magnetic field.
The presence of constant field $B_0$ produces the Zeeman splitting between states $b$ and $d$. As far as energy of states $b,d$  depend on magnetic field linearly, while for states $a,c$ it depends  quadratically, only transition between $b,d$ states take place   in case of a week field. In the following we will consider only transitions between gravitational states in a $1S( b,d)$ manifold.

 Qualitative behavior of the transition probability is given by the Rabbi formula, which can be deduced by neglecting high frequency terms compared to resonance couplings of only two states, initial $i$ and final $f$, in case the field frequency $\omega$ is close to the transition frequency $\omega_{if}=(E_f-E_i)/\hbar$:
\begin{equation}\label{Rabbi}
P=\frac{1}{2}\frac{(V_{if})^2}{(V_{if})^2+\hbar^2(\omega-\omega_{if})^2}\sin^2\left(\frac{\sqrt{(V_{if})^2+\hbar^2(\omega-\omega_{if})^2}}{2\hbar}t\right)\exp(-\Gamma t).
\end{equation}
 The factor $1/2$ appears in front of the right-hand side of the above expression due to the fact that only two $(b,d)$ of four hyperfine states participate in magnetically induced transitions.

It is important that the transition frequencies $\omega_{if}$  do not depend on the antiatom-surface interaction up to the second order in $\Delta$. This is a consequence of the already mentioned fact that all energies of gravitational states acquire equal (complex) shift due to the interaction with a material surface.

\begin{figure}
 \centering
\includegraphics[width=100mm]{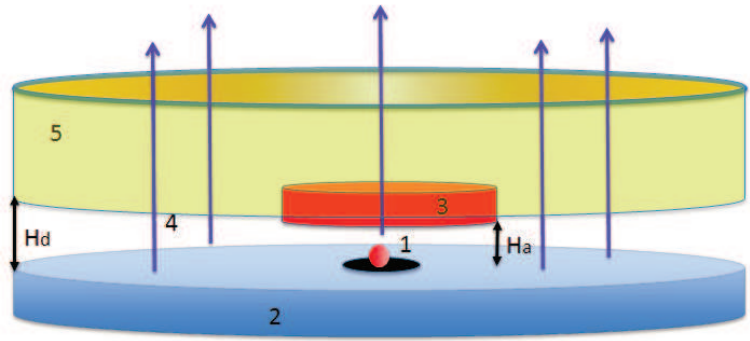}
\caption{A sketch of the principal scheme of an experiment on magnetically induced resonant transitions between $\bar{H}$  gravitational states. 1 - a source of ultracold antihydrogen, 2 -a mirror, 3 -an absorber, 4 -a magnetic field, 5 -a detector.}\label{FigSketch}
\end{figure}

A resonant spectroscopy of $\bar{H}$ gravitational states   could consist in observing   $\bar{H}$ atoms localized in the gravitational field above a material surface at a certain height as a function of the applied magnetic field frequency.
 A "flow-throw type" experiment, analogous to the one, discussed for spectroscopy of neutron gravitational states \cite{ResGranit}, includes  three main steps. A sketch of a principal scheme of an experiment, proposed in \cite{Shaping}, is shown in Fig.\ref{FigSketch}.

%%%%%%%%%%%%%%%%%%%%%%%%%%%%%%%%%%%%%%%%%%%%%%%%%%%%%%%%%%%%%%%%%%%%%%%%
%%%%%%%%%%%%%%%%%%%%%%%%%%%%%%%%%%%%%%%%%%%%%%%%%%%%%%%%%%%%%%%%%%%%%%%%%%%

First, $\bar{H}$ is prepared in a ground gravitational state. This is achieved by passing  $\bar{H}$  through a slit,  formed by a mirror and an absorber placed above it at a given height $H_a$.  A mirror and an absorber form a waveguide with a state dependent transmission  \cite{NeutrWaveGuide}. The choice of $H_a=H_1\simeq 13.6$ $\mu m$ provides that only $\bar{H}$ atoms in a  ground gravitational state pass through the slit.

Second, $\bar{H}$ atoms are affected by alternating magnetic field (\ref{Magn1}) while moving parallel to the mirror. An excited gravitational state is resonantly populated.

Third, the number of $\bar{H}$ atoms in an excited state is measured by means of counting annihilation events in a detector placed at a height $H_d$ above the mirror. The value of $H_d$ is chosen to be  larger, than the spatial size of the gravitational ground state  and smaller than the spatial size of the final state (\ref{Hn}),  $ H_1\ll H_d<H_f$, so that the ground state atoms pass through, while  atoms in the excited state are detected.

%Velocity of ultracold $\bar{H}$ as estimated in \cite{Gbar} is of order of $1$ m/s

\begin{figure}
 \centering
\includegraphics[width=100mm]{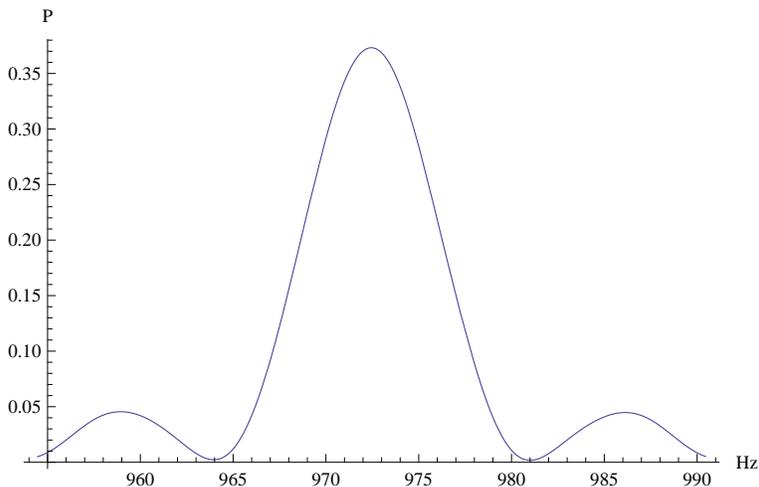}
\caption{Transition probability as a function of the magnetic field frequency for the transition from ground to $6$-th gravitational state.}\label{FigTrans}
\end{figure}

We present a simulation of the number of detected annihilation events as a function of the field frequency in Fig.\ref{FigTrans} for the transition from the ground to the $6$-th excited state, based on numerical solution of equation system Eq.(\ref{Adiab}). Corresponding resonance transition frequency is $\omega=972.46$ Hz. The value of the  field gradient, optimized to obtain the maximum probability of $1\rightarrow 6$ transition during the time of flight $t_{fl}=\tau=0.1$ s,  turned to be equal $\beta=27.2$ Gs/m, the corresponding guiding field value, which guarantees adiabaticity of the magnetic moment motion, is  $B_0=30$ Gs.
%%%%%%%%%%%%%%%%%%%%%%%%%%%%%%%%%%%%%%%%%%%%%%%%%%%%%%%%%%%%%%%%%%%%%%%%
%%%%%%%%%%%%%%%%%%%%%%%%%%%%%%%%%%%%%%%%%%%%%%%%%%%%%%%%%%%%%%%%%%%%%%%%%%%

It follows from (\ref{eGrav}) that the $\bar{H}$ gravitational mass could be deduced from the measured transition frequency $\omega_{nk}$ as follows:
\begin{equation}
M=\sqrt{\frac{2m\hbar \omega_{nk}^3 }{g^2(\lambda_k-\lambda_n)^3}}.
\end{equation}

Let us mention that in the above formula $g$ means the \emph{gravitational field intensity} near the Earth surface, a value which characterizes  properties of field and is assumed to be known with a high precision. At the same time all the information about  gravitational properties of $\bar{H}$ is included in the gravitational mass $M$.

Equality of the gravitational mass $M$ and the inertial mass $m$, imposed by the Equivalence principle, results in the following expression:
\begin{equation}
M=\frac{2\hbar \omega_{nk}^3 }{g^2(\lambda_k-\lambda_n)^3}.
\end{equation}

Assuming that the spectral line width is determined by the lifetime $\tau\approx 0.1$ s of gravitational states, we estimate that the gravitational mass $M$ can be deduced with the relative accuracy $\epsilon_M\sim 10^{-3}$ for $100$ annihilation events for the transition to the $6$-th state.

\section{Conclusion}
We proposed a novel approach to study  gravitational properties of antiatoms based on spectroscopy of quantum states of $\bar{H}$ in the gravitational field of the Earth near a material surface. We showed that the gravitational mass $M$ of the $\bar{H}$ atom can be deduced from measuring the level spacing between gravitational states by means of resonance transitions, induced by alternating gradient magnetic field. We study  main properties of the interaction of gravitationally bound $\bar{H}$  with magnetic field and find the probabilities of resonance transitions from ground to excited states. In the proposed approach the number of annihilation events is countered as a function of the applied field frequency. The spatial positioning of the annihilation detector at a height, corresponding to the classical turning point of the second gravitational state ensures that only $\bar{H}$ atoms in the final  state are detected. The  field gradient required for such a transition, $\beta=27.2$ Gs/m, as well as the  guiding field  $B_0=30$ Gs are easily  accessible in experiments.

The width of the spectral line is determined by the time of life of gravitational states $\tau=0.1$ s. Using transition from the ground to fifth excited state and assuming that the number of detected annihilation events is $100$, the gravitational mass of $\bar{H}$ can be deduced with the precision  $\sim 10^{-3}$. Though systematic effects are assumed to be small, their accurate estimation is underway.
%\bibliographystyle{unsrt}
%\bibliographystyle{prsty}
%\bibliography{hbarclock}

\end{document}